\documentclass[12pt,twoside]{article}
\usepackage{fleqn,espcrc1,epsf}

\usepackage{graphicx}
\usepackage[figuresright]{rotating}

\newcommand{\AmS}{{\protect\the\textfont2
  A\kern-.1667em\lower.5ex\hbox{M}\kern-.125emS}}

\title{Signatures of Quark-Gluon Plasma Phase Transition in High-Energy
     Nuclear Collisions}

\author{Cheuk-Yin Wong \address{Physics Division, Oak Ridge National
        Laboratory$\dagger$,\\Oak Ridge, TN 37831, U.S.A.}%
        \thanks{$^{\dagger}$Managed by UT-Battelle, LLC for the
        U.S. Department of Energy under Contract 
        DE-AC05-00OR22725.}}

\begin{document}

\maketitle

\begin{abstract}
In high-energy nuclear collisions, the new phase of the quark-gluon
plasma is indicated by an anomalous increase in pressure, an excess of
direct photon production, an excess of strangeness production, and an
anomalous $J/\psi$ suppression.  We review these signatures and
discuss how recent high-energy heavy-ion experiments at CERN are
consistent with the production of the quark-gluon plasma in
high-energy Pb+Pb collisions.
\end{abstract}

\section{INTRODUCTION}

A recent news release from CERN stated that a new state of matter, the
quark-gluon plasma (QGP), was created by high-energy heavy-ion
collisions at CERN.  The evidence for  such an observation was
summarized by Heinz and Jacobs \cite{Hei00}.  The news release has
generated a great deal of excitement. It is useful to review and
re-examine this evidence from various perspectives to see whether such
an excitement is justified or not.

In this review and re-examination, we shall discuss the nature of the
order of the hadron-QGP phase transition, the dynamics of  how such a
transition takes place in high-energy heavy-ion collisions, and the
signatures for the phase transition.  Prominent signatures include the
large pressure in the quark-gluon plasma phase which leads to an
anomalous increase in the freeze-out volume, the emission of direct
photons when the matter is in the quark-gluon plasma phase
\cite{Fei76}, the strong enhancement of strangeness \cite{Raf82}, and
the suppression of $J/\psi$ production \cite{Mat86}.  Other evidence
can also be found by mapping out the phase boundary in the phase
diagram from the distribution and the yield of hadron products, which
will be reported by Braun-Munzinger in these Proceedings \cite{Bra00}.
Our re-assessment of these signatures provides additional support to
indicate that, in line with the summary of Heinz and Jacobs
\cite{Hei00}, the experimental data appear to be consistent with
the occurrence of the quark-gluon plasma.

\section{THE QUARK-GLUON PLASMA PHASE TRANSITION}

A phase transition is classified as a first- or second-order
transition depending on how its free energy varies with the order
parameter.  The phase transition is first order if the free energy as
a function of the order parameter exhibits two local minima at
different values of the order parameter, and the order parameter of
the lowest minimum changes from one local minimum to the other
discontinuously as the temperature passes the critical temperature
$T_{\rm crit}$ as shown schematically in Fig.\ 1$a$.  A phase
transition is second order if the free energy as a function of the
order parameter exhibits a single minimum, and the location for the
minimum changes continuously for different values of the order
parameter as the temperature passes the critical temperature $T_{\rm
crit}$ (Fig.\ 1$b$).  (See also Fig.\ 11.3 and 11.4 of Ref.\
\cite{Won94}.)

\null\vskip 6.0cm \epsfxsize=250pt
\includegraphics{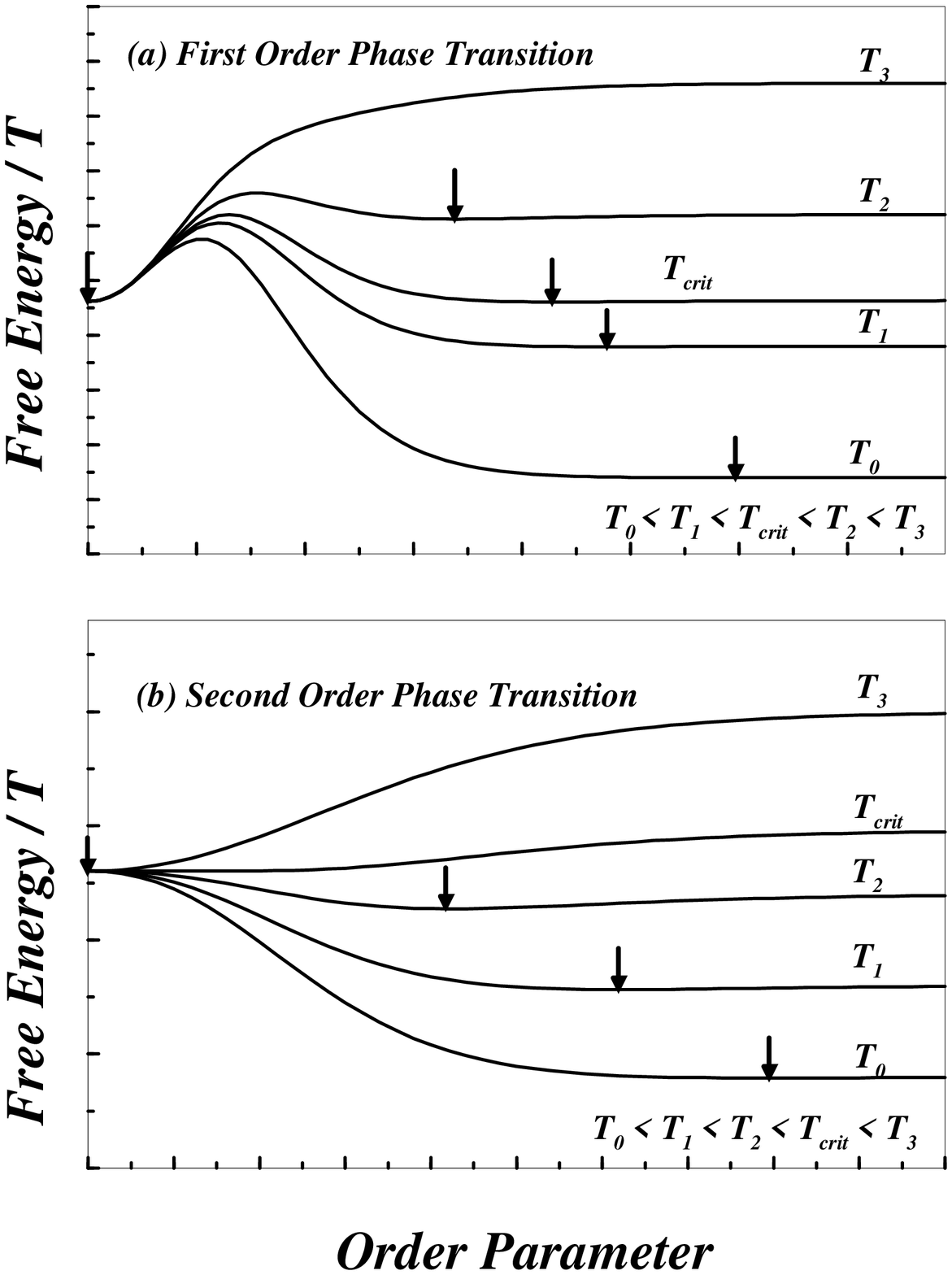}
\vskip 3.5cm
\null\hskip 6.8cm
\begin{minipage}[t]{8.7cm}
\noindent
\bf Fig.\ 1 { \rm { The free energy divided by the temperature as a
function of the order parameter for different temperature $T$.  Fig.\
1$a$ shows a first-order phase transition, and Fig.\ 1$b$ shows a
second-order phase transition.  An arrow indicates the location of
a local minimum.}}
\end{minipage}
\vskip 4truemm
\noindent 

\vskip -5.4in \hangafter=-30 \hangindent=-3.6in The order parameter
for strongly\break interacting matter is the Wilson line parameter,
which is related to the interaction energy $V$ when an isolated quark
is placed in the medium.  It is proportional to $e^{-Vt}$ for a Wilson
line of temporal length $t$.  In the hadron phase, because of the
linear interaction between a quark and an antiquark, the interaction
energy $V$ is infinite when an isolated quark is placed in the\break
medium and the Wilson line parameter is zero.  The Wilson line
parameter is non-zero in the deconfined phase.

\hangafter=-30 \hangindent=-3.6in Lattice gauge calculations show that
the transition is first order for a quark-gluon plasma with two
flavors, but\break changes to a second-order phase transition for
three flavors with the physical mass of the strange quark
\cite{Aok99}.\break  These theoretical predictions need to confront
experimental data to deduce the nature of the phase transition.

\hangafter=-4 \hangindent=-3.6in How does one visualize the difference
between the hadron matter and the quark-gluon plasma?  One can
envisage a spatial distribution of the link variables which contain
the gluon degrees of freedom and the generators of SU(3).  The link
variables in strongly interacting matter are the analogue of spins in
a spin gas.  The hadron phase is the low-temperature phase in which
neighboring link variables (in a plaquette) are all correlated in
order to reside in the lowest energy state.  The correlation can be
considered roughly as some generalized `alignment' of the link
variables, in analogy with the alignment of spins in the spin
lattice gas.  In such an `aligned' state of the link variables in all
regions of space, the interaction between a quark and an antiquark at
large distances becomes a linear function of the distance, which leads
to the confinement of quarks and gluons.  In contrast, in the
high-temperature quark-gluon plasma phase, the temperature is so high
that the tendency to align the link variables due to the QCD
interaction is overwhelmed by the tendency to disalign the link
variables due to the thermal fluctuation.  The link variables are no
longer aligned.  The interaction between a quark and an antiquark
above the critical temperature at large distances is no longer
governed by the linear interaction, and the quarks and gluons become
deconfined.

The deconfinement phase is further accompanied by the restoration of
chiral symmetry~\cite{Kar99} in which the light quarks in the hadron
phase are restored to their nearly massless current quark mass values.
Thus, the phase transition from hadron matter to the quark-gluon
plasma can be described as a change of the link variables from an
ordered aligned state to a state of disorientation, accompanied by an
abundant production of nearly massless quark pairs and additional
gluons.  As the link variables representing the gluons are now free to
orient randomly and the nearly massless light quark pairs are
copiously produced, the degree of freedom is greatly enlarged and the
energy density of the deconfined phase is much greater as a result.

\section{HEAVY-ION COLLISIONS TO PRODUCE THE PHASE TRANSITION}

\hangafter=2 \hangindent=-8.2cm High-energy heavy-ion collisions can
be used to produce a phase transition from the hadron matter to the
quark-gluon plasma.  Because of the Lorentz contraction, the
nucleon-nucleon collisions in a nucleus-nucleus collision occur
at about the same time and at nearly the same spatial proximity.  As a
consequence, a region of very high energy density can be created.

\null\vskip -3.1cm \epsfxsize=250pt
\includegraphics{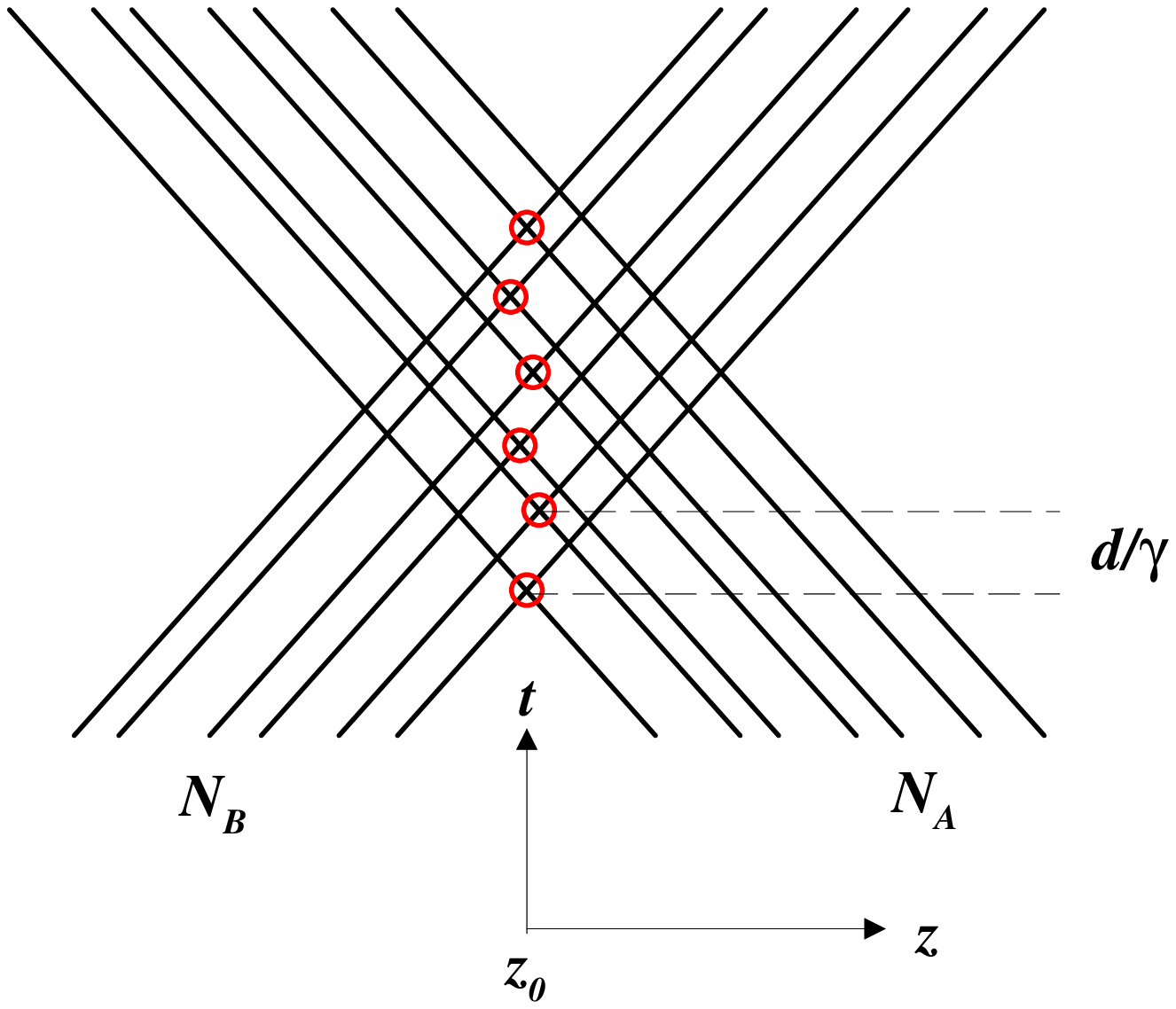}
\vskip  6.5cm
\null\hskip 7.8cm
\begin{minipage}[t]{7.5cm}
\noindent
\bf Fig.\ 2. { \rm { The space-time diagram of a row of $N_B$ projectile
nucleons colliding with a row of $N_A$ target nucleons.}}
\end{minipage}
\vskip 4truemm
\noindent 

\vskip -6.2cm \hangafter=-20 \hangindent=-8.2cm The dynamics of such a
nucleus-nucleus collision can be best described in the nucleon-nucleon
center-of-mass system.  For a given impact parameter, we can divide
the\break transverse area of the colliding nuclei into rows with an
area of a nucleon-nucleon cross section $\sigma_{in}$.  Within each
row, each projectile nucleon will collide with each target nucleon.
By assuming straight-line space-time trajectories, the space-time
diagram\break of these nucleons is shown in Fig.\ 2.

\hangafter=-1 \hangindent=-8.5cm We can focus our attention at one
spatial point at $z_0$ in this row.  A nucleon-nucleon collision at
this point (represented by an open circle in Fig.\ 2) will lead to the
deposition of energy about that point.  At a proper time of about 1
fm/c, the deposited energy will materialize as field quanta of matter
in the form of hadrons if the matter is favored to be in the hadronic
state.  The field quanta will be quarks and gluons if the matter is
favored to be in the quark-gluon plasma state.

There are many other nucleon-nucleon collisions which take place
sequentially at $z_0$ at a time interval of $\sim d/\gamma$, as
represented by the open circles in Fig.\ 2.  Here, $d\sim 2.5 $ fm is
the average nucleon-nucleon spacing in nuclear matter at rest and
$\gamma$ is the relativistic factor for the motion of the nuclei in
the nucleon-nucleon center-of-mass system.  Each later nucleon-nucleon
collision at the same collision point deposits additional energy.  The
local energy density increases as a function of time and is
approximately proportional to the number of nucleon-nucleon collisions
$N$ occurring at that point:
\begin{eqnarray}
\epsilon \sim  N ~{dn \over dy} {m_t \over \sigma_{in} d/\gamma},
\end{eqnarray}
where $dn/dy$ is the average multiplicity per unit of rapidity at
$y_{{}_{CM}}$$=$0 for a nucleon-nucleon collision, and $m_t$$\sim$$0.35$
GeV is the transverse mass of a produced pion.  For collisions at 158A
GeV per nucleon, the energy density deposited per nucleon-nucleon
collision as given by Eq. (1) is approximately $N\times$1 GeV/fm$^3$.

In actual nucleus-nucleus collisions, nucleons lose energy as they
collide and the multiplicity distribution in later collisions will be
slightly lower than those from earlier collisions.  Equation (1),
which is defined in terms of an average multiplicity value, is a
simplifying, but useful, relation which gives an approximate estimate
of the local energy density.

As the size of the colliding nuclei increases, the energy density
deposited at a collision point in a central collision will increase.
When the number of nucleon-nucleon collisions at a spatial point
exceeds a critical number $N_c$, the local energy density will
increase beyond the critical energy density $\epsilon_c$ for
transition to the quark-gluon plasma phase \cite{Won96}.  Then, the
state of lowest free energy becomes the QGP phase with field quanta of
quarks, antiquarks, and gluons.  If the number of nucleon-nucleon
collisions at a point is lower than the critical collision number $N_c$
, it will not reach the critical energy density and the field quanta
will be hadrons consisting mostly of pions.

\section{ANOMALOUS FREEZE-OUT VOLUME}

In a first-order phase transition, the energy density and pressure
change abruptly from the hadron phase to the quark-gluon plasma phase.
Consequently, the pressure changes abruptly as a function of the local
collision number when the collision number passes the critical value,
$N_c$.

The pressure in the quark-gluon plasma is given approximately by
$P_{\rm QGP} \sim37 \pi^2T^4/90 $ which is much greater than the
pressure of a pion gas, $P_{\rm pion}=6 \pi^2T^4/90 $.  Thus, the presence
of a quark-gluon plasma is characterized by a region of very high
pressure.  The large pressure of the quark-gluon plasma can be
detected by a large freeze-out volume.

We would like to examine the relation between the initial pressure and
the freeze-out volume in Bjorken hydrodynamics \cite{Bjo83}.  The
matter produced in high-energy heavy-ion collisions is subject to a
strong longitudinal expansion.  We consider a plateau distribution of
matter within rapidity range $|y|<y_0$ and longitudinal coordinates
$|z|<|z_0|$.  The initial proper time is then $\tau_0=|z_0|/\sinh
y_0$.  The solution for the pressure $P$ in Bjorken hydrodynamics is
\begin{eqnarray}
P(\tau,y)=P(\tau)\theta(y_0-|y|),
\end{eqnarray}
where
\begin{eqnarray}
{P(\tau)\over P(\tau_0)} = \left ({ \tau_0 \over \tau} \right )^{4/3} .
\end{eqnarray}
(See Eq.\ (13.17$b$) of Ref.\ \cite{Won94}.)  At the proper time
$\tau$, the boundary of the matter is expanded up to $z_\tau =\pm\tau \sinh
y_0$, and thus the ratio of $|z_\tau|/|z_0|$ is
\begin{eqnarray}
{|z_\tau| \over |z_0|} = \left ( {P(\tau_0)\over P(\tau)} \right )^{3/4}.
\end{eqnarray}
As the matter expands, there will be a pressure $P(\tau_f)$, the
freeze-out pressure, at which particles will no longer interact and
will exhibit freeze-out characteristics.  Because the volume of matter
$V$ is proportional to $|z_\tau|$, the ratio of the freeze-out volume
$V(\tau_f)$ at $\tau_f$ to the initial volume $V(\tau_0)$ is
\begin{eqnarray}
\label{eq:PPP}
{V(\tau_f) \over V(\tau_0) } = \left ( {P(\tau_0)\over P(\tau_f)} \right
)^{3/4}.
\end{eqnarray}
Hence, the freeze-out volume $V(\tau_f)$ varies with the initial
pressure as $[P(\tau_0)]^{3/4}$. 

Equation (\ref{eq:PPP}) is applicable to hadron matter.  It is also
applicable approximately to the QGP undergoing strong longitudinal
expansion, for which the temperature of the QGP is lowered so rapidly
that the expansion drives the system below the critical temperature.
When the energy density and pressure of the supercooled quark-gluon
plasma drop down to the same level as those of the hadron matter,
spontaneous transition to hadron matter will take place at proper time
$\tau_H$ resulting in hadron matter of volume $V(\tau_H)$.  The hadron
matter then undergoes further longitudinal expansion to reach the
freeze-out volume $V(\tau_f)$ at proper time $\tau_f$.  For the
quark-gluon plasma subject to a strong longitudinal expansion, we
have
\begin{eqnarray}
{V(\tau_f) \over V(\tau_0) } =
{V(\tau_f) \over V(\tau_H) }{V(\tau_H) \over V(\tau_0) }
\sim 
 \left ( {P(\tau_H)\over P(\tau_f)} \right)^{3/4}  \left ( {P(\tau_0)\over
P(\tau_H)} \right )^{3/4}
=
 \left ( {P(\tau_0)\over P(\tau_f)} \right
)^{3/4}.
\end{eqnarray}
The freeze-out volume varies approximately with the initial QGP
pressure as $[P(\tau_0)]^{3/4}$.

Based on the above, the large pressure of the quark-gluon plasma will
lead to a large freeze-out volume and a first-order phase transition
will be indicated by an anomalous increase of the freeze-out volume
due to the increase in the pressure in the QGP phase.  We expect that
the anomalous increase in the freeze-out volume should occur when the
average number of nucleon-nucleon collisions passes through the
critical value $N_c$.  The average nucleon-nucleon collision number in
a collision is a function of centrality or multiplicity, so the
freeze-out volume should increase anomalously as a function of
centrality or multiplicity in a first-order phase transition.  In
contrast, the increase  should be much more gradual in a second-order
phase transition.

In actual heavy-ion collisions, the freeze-out volume also increases
with centrality because of the increase in the volume of the
overlapping region. Consequently, there will be the systematics of
such a normal increase of the freeze-out volume, as one goes from the
peripheral region to the central region.  However, the increase in the
freeze-out volume due to the large pressure of the quark-gluon plasma
is an anomalous addition to these systematics.  By looking at the
systematics of the freeze-out volume as a function of centrality or
multiplicity, one should be able to separate out these two components:
the normal component due to the increase in the volume of the
collision region for peripheral and semi-central collisions, and an
anomalous part due to the high pressure of the quark-gluon plasma
which becomes important for central collisions.

\hangafter=5 \hangindent=-9.1cm Recently the NA44 Collaboration
\cite{NA4400} observed that the freeze-out volume increased
anomalously from semi-central collisions to central collisions, for
Pb+Pb collisions at 158A GeV (Fig.\ 3).  This interesting observation
by the NA44 Collaboration may be evidence for the quark-gluon plasma.
In semi-central collisions, the producded matter are likely to be
hadron matter.  Therefore, the expansion and the freeze-out volume is
governed mainly by the expanding hadron matter with $P(\tau_0) \sim
P_{\rm pion}$ and $V_{\rm freeze-out} \propto P_{\rm pion}^{3/4}$.
For central collisions, if quark-gluon plasma is produced, the
expansion and cooling for these central collisions is governed
essentially by the initial pressure of the quark-gluon plasma,
$P(\tau_0)$$\sim$$P_{QGP}$,\break which is a very high pressure,
and\break $V_{\rm freeze-out} \propto P_{\rm QGP}^{3/4}$.  Therefore,
this expansion will lead to an anomalous increase in the freeze-out
volume for central collisions, much greater than what one would expect
from the systematics for semi-central collisions.  The rapid increases
of the freeze-out volume as a function of multiplicity, as observed by
the NA44 Collaboration, may indicate that the hadron-QGP phase
transition is a first-order phase transition.  Further investigations
are needed to shed more light on this interesting observation of the
NA44 Collaboration.

\null\vspace*{-5.9cm} 
\epsfxsize=250pt
\includegraphics{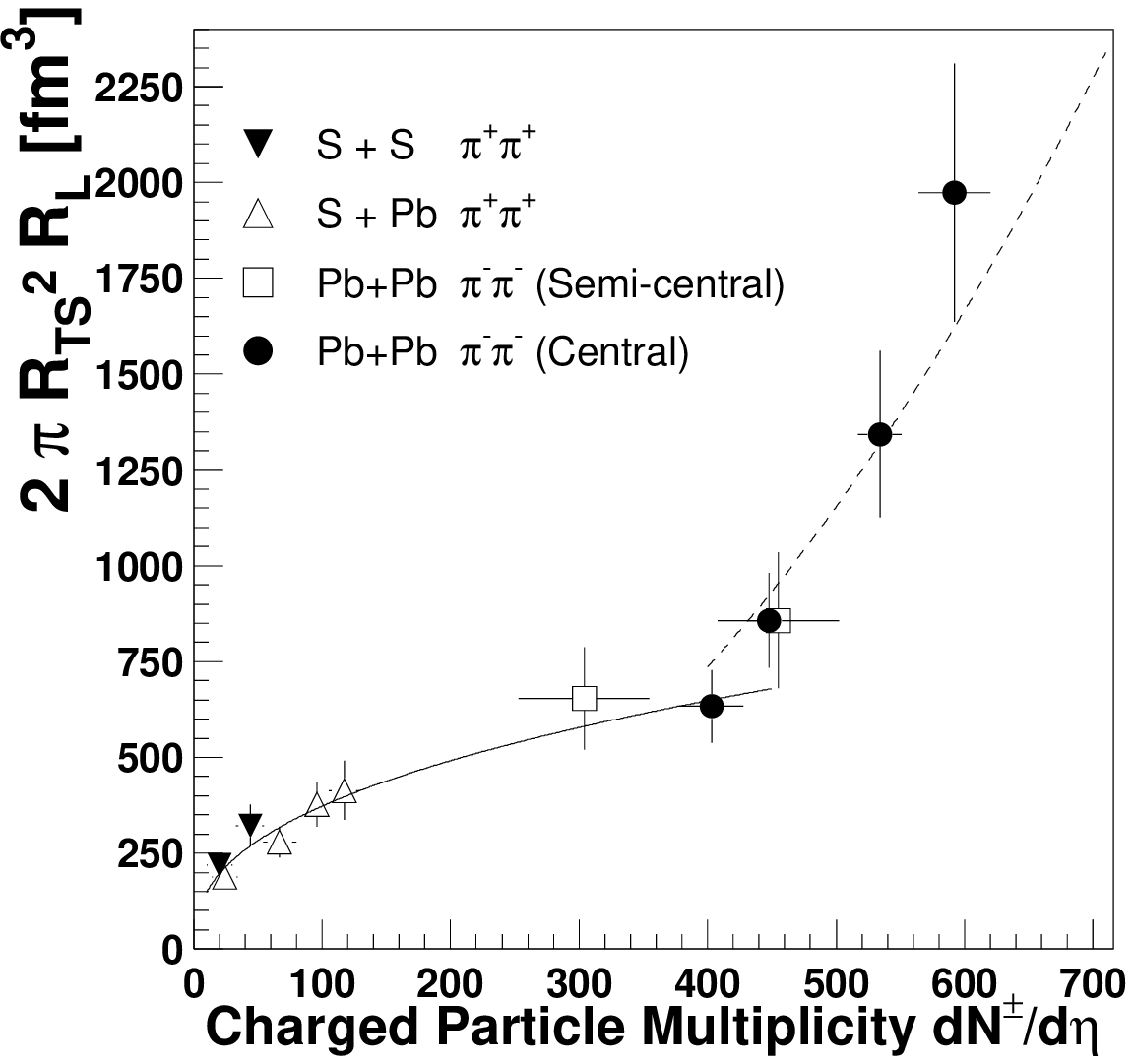}
\vspace*{ 1.7cm}
\null\hskip 7.0cm
\begin{minipage}[t]{8.4cm}
\noindent
\bf Fig.\ 3.\ { \rm { The NA44 data of the freeze-out volume as a
function of $dN^{\pm}/d\eta$ \cite{NA4400}.  The curves joining the
data points are parametrizations in the form of
$(dN^{\pm}/d\eta)^\alpha$ from the NA44 Collaboration.}}
\end{minipage}
\vskip 4truemm
\noindent

\vspace*{0.8cm}
\section{DIRECT PHOTON PRODUCTION} 

As the quark-gluon plasma expands, its energy density and temperature
decreases.  At the point when its temperature decreases below the
critical temperature for hadron-QGP phase transition, matter will
undergo a phase transition from the quark-gluon plasma to the hadron
phase.

During the time when the matter is in the quark-gluon plasma phase, it
will emit particles.  Photons arising from the electromagnetic
interactions of the constituents of the plasma will provide
information on the properties of the plasma at the time of their
emission.  Since photons are hardly absorbed by the medium, they form
a relatively `clean' probe to study the state of the quark-gluon
plasma.  The presence of these photons in high-energy heavy-ion
collisions can also possibly provide evidence for the production of the
quark-gluon plasma \cite{Fei76,Kaj86,Kap91}.

Photons are also produced by many other processes in heavy-ion
reactions. They can come from the decay of $\pi^0$ and $\eta^0$.  As
$\pi^0$ particles are copiously produced in strong interactions
between nucleons, photons coming from the decay of $\pi^0$ are much
more abundant than photons produced by electromagnetic interactions of
the constituents of the quark-gluon plasma.  The photons from the
decay of $\pi^0$ and $\eta^0$ can be subtracted out by making a direct
measurement of their yields, obtained by combining pairs of photons.
Because of the large number of $\pi^0$ produced, this subtraction is a
laborious task, but much progress has been made to provide meaningful
results after the subtraction of the photons from the $\pi^0$ and
$\eta^0$ backgrounds \cite{WA9800}.  Photon measurements obtained
after the subtraction of the photons from meson decays are
conveniently called measurements of ``direct photons''.

\vspace*{0.2cm}
\hangafter=2 \hangindent=-3.7in 
Direct photons are produced from the interaction of matter in the QGP
phase, a mixed QGP and hadron phase, a pure hadron gas, and hard QCD
processes.  Different processes give rise to photons
in different momentum regions.  One may wish to go to the region of
photon transverse momentum $p_{T}>$ 2 GeV/c to minimize the
contributions from hadrons.  If a hot quark-gluon plasma is formed
initially, clear signals of photons from the plasma could be visible
by examining photons with $p_{T}$ in the range 2$-$3 GeV/c
\cite{Kaj86,Kap91}.  On the other hand, photons in this region of
transverse momenta are also produced by the collision of partons of
the projectile nucleons with partons of the target nucleons.  Such a
contribution must be subtracted in order to infer the net photons from
the quark-gluon plasma sources.

\vspace{ -10.5cm} \epsfxsize=250pt
\includegraphics{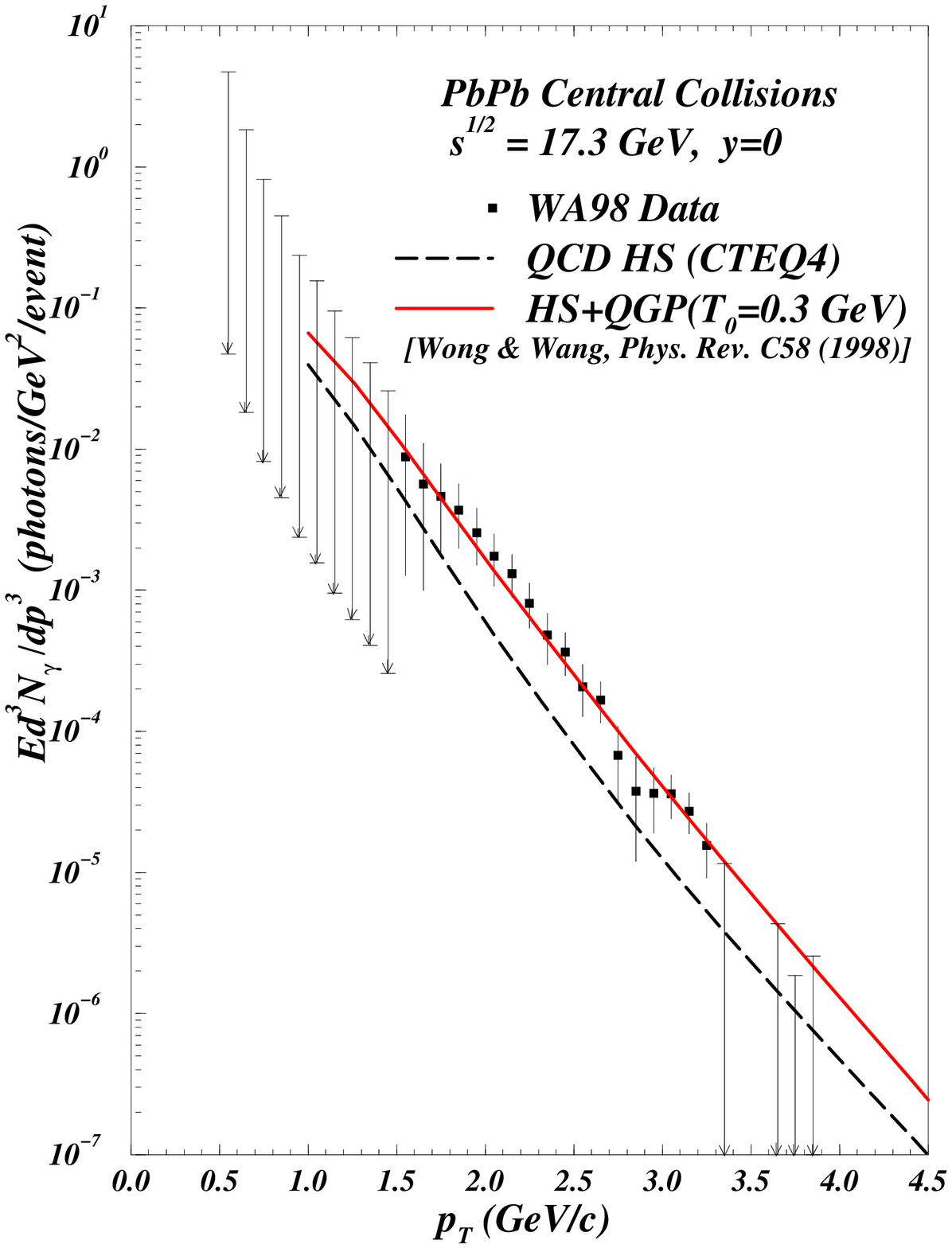}
\vspace{  11.cm}
\null\hskip 6.8cm
\begin{minipage}[t]{8.7cm}
\noindent
\bf Fig.\ 4. { \rm { The spectrum of direct photons.  The data are from
the WA98 Collaboration \cite{WA9800}, and the curves are from Ref.\
\cite{Won98}.}}
\end{minipage}
\vskip 4truemm
\noindent 

\vskip -2.8cm
\hangafter=-5 \hangindent=-3.7in 
Recently, the WA98 Collaboration has
measured the photon spectrum for central Pb+Pb collisions at 158A GeV
\cite{WA9800}.  The photon spectrum from the hard scattering of the
nucleons has been calculated previously by taking into account the
next-to-leading order contributions and the intrinsic transverse
momentum of partons.  The results \cite{Won98} scaled to the Pb+Pb
collisions are shown as the dashed curve in Fig.\ 4.  One finds that
the WA98 experimental data is in excess of the contributions from the
hard scattering of the nucleons.

We can calculate the total photon spectrum including the contributions
from the nucleon hard scattering and the quark-gluon plasma.  If we
assume a quark-gluon plasma with a transverse area of the overlapping
area appropriate for the corresponding impact parameter and an initial
plasma formation time at $\tau$$=$1 fm/c, with a temperature of 300
MeV cooling to a critical temperature of 200 MeV, we obtain the photon
spectrum in Fig.\ 9 of Ref.\ \cite{Won98} shown here as the solid
curve.  The good agreement of the model with the WA98 data provides
additional support for the production of the quark-gluon plasma in
high-energy central Pb+Pb collisions.

\section{Strangeness enhancement}

The strangeness content is enhanced in hadron matter as the
temperature increases, but the strangeness is enhanced to an even
greater extent in a quark-gluon plasma.  The greater strangeness
enhancement arises from a higher temperature in the quark-gluon plasma
and from its lower effective light quark masses because of the
restoration of chiral symmetry. As the strangeness content is greatly
enhanced, the probability for the production of multi-strange hyperons
will also be greatly enhanced in a quark-gluon plasma \cite{Raf82}.

In a recent measurement of the WA97 Collaboration \cite{Wa9799}, the
production of multi-strange hyperons is found to be substantially
enhanced.  In particular, the production of $\Omega^-+\overline {
\Omega^+}$ in Pb+Pb collisions at 158A GeV is enhanced by up to a
factor of 15 relative to that of p+Be.  A more detailed description of
the strangeness enhancement will be presented by Odyniec in these
Proceedings \cite{Ody00}.

Multi-strange hyperons can also be produced by secondary collisions of
hadrons.  It is known that the collision of the produced pions with
nucleons leads to the enhancement of kaons and $\Lambda$ particles.
Repeated collisions of kaons in the medium with a $\Lambda$ particle
can raise the strangeness of the hyperon by one or more units. The
enhancement of hyperons by these secondary collisions increases with
the increase in the size of the colliding nuclei, and it is important
to take this increase into account.  Recent comparison of the hyperon
yields as obtained by theoretical RQMD cascade model calculations
\cite{Sor95} shows that the RQMD calculations can reproduce the yields
of strange particles with one or two units of strangeness, but it
underpredicts the yield of $\Omega^-$ by a factor of 3, and
underpredicts the yield of $\overline {\Omega^+}$ by about 40 \%
\cite{Wa9799a}.

The discrepancies of the yields of the $\Omega$ hyperons with the RQMD
cascade model may be additional evidence for the production of the
quark-gluon plasma in Pb+Pb collisions.  This may be the case, but such
a conclusion relies heavily on the assumed input of many
strangeness-raising cross sections, for which no experimental data are
available.  The evidence will be substantially strengthened if the
strangeness-raising cross sections can be better determined by a
reliable theoretical model, such as the quark-interchange model of
Barnes and Swanson \cite{Bar92}.  As emphasized by Ko \cite{Ko00}, the
strangeness-raising cross section for the reaction such as $K^0 + \Xi^-
\to \pi^0 + \Omega^-$ can proceed through the interchange of a
strange quark from $K$ to $\Xi$, which need not be OZI-suppressed as
in the $\pi + N \to K + \Lambda$ reaction, where an intermediate
strange quark pair is produced.  Future theoretical evaluation of
these strangeness-raising cross sections will be of great interest in
clarifying the origin of the enhancement of the $\Omega$ hyperons in
Pb+Pb collisions.

\section{$J/\psi$ Suppression}

In a quark-gluon plasma the screening of the charm quark and antiquark
will make the $J/\psi$ unbound, and its production will be suppressed.
The occurrence of $J/\psi$ suppression has been suggested as a
signature for the quark-gluon plasma \cite{Mat86}.  The experimental
observation by the NA50 Collaboration \cite{Gon96,Rom98,NA5000} of an
anomalous $J/\psi$ suppression in Pb+Pb collisions has led to the
suggestion that this anomalous suppression arises from the production
of the quark-gluon plasma {\cite{Won96,Kha96,Bla96}.  The phenomenon
of anomalous $J/\psi$ suppression has also been studied by many
authors \cite{Gav96}.

The production of $J/\psi$ is suppressed not only by the quark-gluon
plasma but also by the collision of the $J/\psi$ (or its precursor)
with nucleons and produced particles.  The absorption of $J/\psi$ by
these particles was considered in a simple analytical model
\cite{Won96}.  One divides the transverse area into rows of nucleons
of cross section $\sigma_{in}$.  The projectile nucleons in each row
collide with the target nucleons within the same row. One assumes
simple straight-line space-time trajectories for the colliding
nucleons as in Fig.\ 2.  In this space-time diagram, each
nucleon-nucleon collision is a possible source of $J/\psi$ production
and also a source of produced particles which will absorb $J/\psi$.
By parametrizing the dissociation cross section of $J/\psi$ in its
collision with nucleons or produced particles, the $J/\psi$ production
cross section in a nucleus-nucleus collision can be evaluated.

It was found that the simple model of $J/\psi$ absorption by nucleons
and produced particles can explain the $p$+A, O+A, and S+U data, but
not the Pb+Pb data.  The deviation of Pb+Pb data from the theoretical
extrapolations suggests that there is a transition to a new phase of
strong absorption, the quark-gluon plasma, which sets in when the
local energy density exceeds a certain threshold.  As we remarked
before, the energy density at a particular spatial point is
approximately proportional to the number of nucleon-nucleon collisions
which have taken place at that point.  We postulate that the matter at
a point makes a transition to the new phase of strong $J/\psi$
absorption if there have been $N_c$ or more nucleon-nucleon collisions
at that point.

\null\vskip 10.3cm \epsfxsize=250pt
\includegraphics{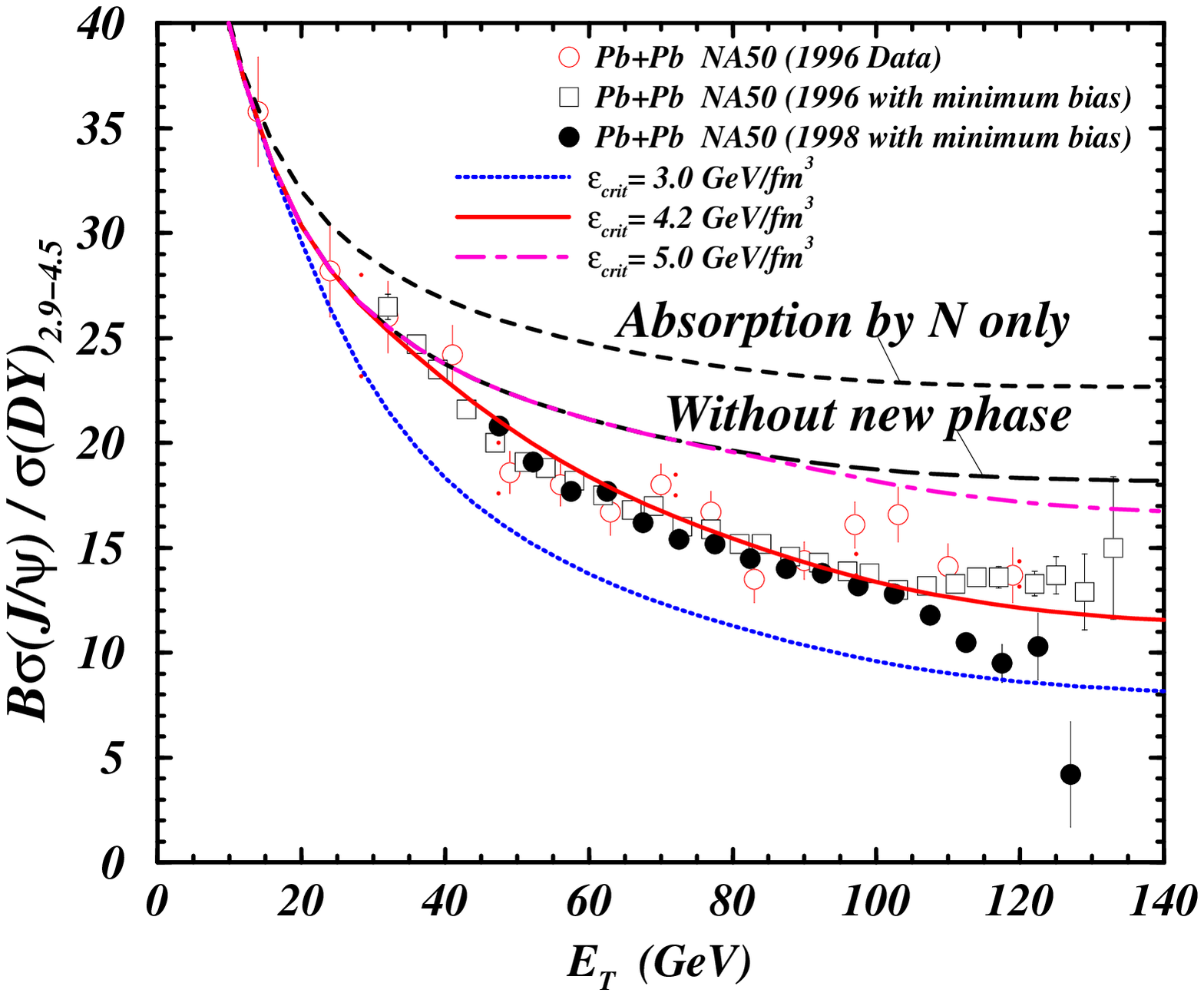}
\vskip -3.8cm
\null\hskip 6.8cm
\begin{minipage}[t]{8.7cm}
\noindent
{\bf Fig.\ 5.} { The ratio of the $J/\psi$ cross section to the
Drell-Yan cross section at different values of transverse energy
$E_T$.  The NA50 data are from Ref.\ \cite{Gon96,Rom98,NA5000}.}
\end{minipage}
\vskip 4truemm
\noindent 

\vskip -9.7cm \hangafter=-30 \hangindent=-3.7in We assume that the
deconfinement temperature is greater than the dissociation
temperatures for the dissociation of both $\chi$ and $J/\psi$.  Then,
$N_c$$=$4 gives very good agreement with the pA, O+A, S+U, and Pb+Pb
data.  The critical nucleon-nucleon collision number, $N_c$$=$4,
corresponds to a critical energy density of $\epsilon_c$$=$4.2
GeV/fm$^3$. The results for the ratio of the $J/\psi$ cross section to
the Drell-Yan cross section is shown as the solid curve in Fig.\ 5.
The good agreement of this critical energy density model \cite{Won96}
with the WA50 data provides evidence for the production of the
quark-gluon\break  plasma in central high-energy Pb+Pb collisions.

\hangafter=-1 \hangindent=-3.7in 
The analytical model of Ref.\ \cite{Won96} for $J/\psi$ is useful
to provide new insight into the main characteristics of the
suppression process.  It is desirable to make a more refined
calculation to include many improvements in order to see whether the
main characteristics may depend on these refinements.  The cross
section for the dissociation of $J/\psi$ by hadrons should be reliably
calculated and their dependence on hadron types and hadron energies
included.  The population of the $\rho$ meson should also be obtained
in a dynamical model where the $\rho$ mesons are both formed by pion
collisions and decay back into pions.  The effect of the degradation
of the energy of the colliding nucleons and their subsequent lower
hadron production should also be taken into account.  These
improvements can be included in a Monte Carlo cascade model
calculation as shown below.

To obtain reliable $J/\psi$ dissociation cross sections, we first
determine the effective quark-gluon potential from hadron masses and
use these interactions in the quark-interchange model of Barnes and
Swanson \cite{Bar92} to obtain the dissociation cross section.  The
Barnes and Swanson model has been tested and found to give good
agreement with experimental phase shifts and cross sections in many
meson-meson reactions.

By using the Barnes and Swanson model, the cross section for the
dissociation of $J/\psi$ by a pion is found to be small, with a peak
cross section of about 1 mb above the threshold of 640 MeV.  On the
other hand, the cross section for the dissociation by a $\rho$ meson
is large because it is above the $D+\bar D$ threshold.  The cross
section decreases rapidly as a function of energy. The dependence of
the cross sections on the mass of the $\rho$ meson have also been
calculated \cite{Won99}.

\hangafter=2 \hangindent=-3.7in Having obtained the dissociation cross
sections, we incorporate them in a heavy-ion Monte Carlo cascade
program to study the survival probability of a produced $J/\psi$.  We
use different basic numerical programs in our Monte Carlo cascade
calculations \cite{Won00a}.  In one of our programs for which we shall
report our results, we base our calculations on the MARCO program,
which is a multiple collision model with a stopping law to govern the
degradation of the energy of nucleons and a particle production law to
specify the distribution of produced particles after a\break
nucleon-nucleon collision \cite{Won89}.  The MARCO program has been
found to give a good description of the gross features of hadron
production \cite{Zha92}.\break The rescattering of the produced
particle is now incorporated to follow the cascade of the hadrons.

\null\vspace*{2.3cm}
\epsfxsize=250pt
\includegraphics{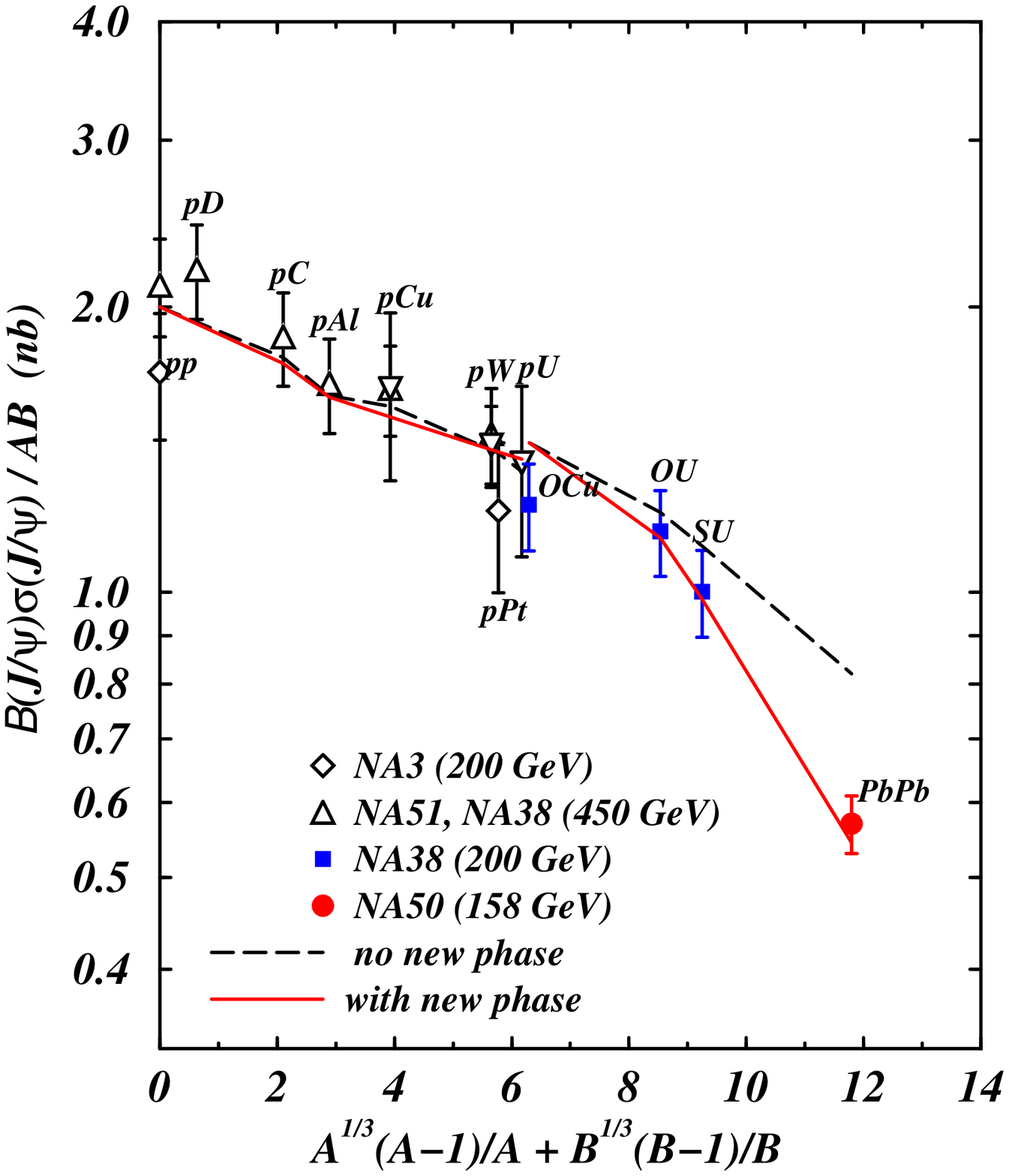}
\vskip -1.1cm
\null\hskip 6.8cm
\begin{minipage}[t]{8.7cm}
\noindent
\bf Fig.\ 6. { \rm
{$J/\psi$ cross section as a function of $A^{1/3}(A-1)/A+B^{1/3}(B-1)/B$.
\cite{NA4400}.}}
\end{minipage}
\vskip 4truemm
\noindent 

\vskip -3.5cm \hangafter=-7 \hangindent=-3.7in In Fig.\ 6, we plot the
$J/\psi$ production cross section as a function of
$A^{1/3}(A-1)/A+B^{1/3}(B-1)/B$, where $A$ and $B$ are the mass numbers
of the colliding nuclei.  The results from the Monte Carlo cascade
calculations are given as the dashed curve.  The inclusion of the
$\rho$ mesons increases the absorption slightly for S+U and Pb+Pb but
does not lead to good agreement with the Pb+Pb data. Again, when
we postulate that a new phase of $J/\psi$ absorption is produced when
the local energy density exceeds a critical value, then we obtain the
results as shown in the solid curve.  The qualitative features of the
$J/\psi$ absorption are not changed with a refined calculation for
$J/\psi$ production.

\section{CONCLUSION AND SUMMARY}

In high-energy heavy-ion collisions, the occurrence of the quark-gluon
plasma will be accompanied by an anomalous increase in pressure, an
excess of direct photon production, an excess of strangeness
production, and an anomalous $J/\psi$ suppression.  We reviewed these
signatures and find that they are present in central Pb+Pb collisions
at 158A GeV.

For the indicator of the anomalous increase in pressure, the NA44 data
exhibits an anomalous increase in the freeze-out volume as a function
of multiplicity.  Superimposing on the normal systematic increase of
the freeze-out volume due to the increase in the volume of the
interacting region may be the additional increase due to the high pressure
of the quark-gluon plasma.  

For the question of direct photon production, the WA98 Collaboration
has measured the direct photon spectrum in central Pb+Pb collision,
and has found that the photon yield is in excess of the expected yield
from nucleon-nucleon hard scattering processes.  The excess direct
photons can be understood as arising from a quark-gluon plasma with an
initial temperature of 300 MeV.

For the enhancement of strangeness, the WA97 Collaboration observed a
large enhancement in the yield of $\Omega$ hyperons which cannot be
explained by RQMD cascade calculations.  Such a discrepancy may
support the enhancement as originating from the quark-gluon plasma.
This evidence needs to be confirmed by calculating the
strangeness-raising cross sections with a reliable model.

In the case of $J/\psi$ suppression, the anomalous suppression in
central Pb+Pb collisions cannot be explained by the absorption by
hadrons.  This was demonstrated in a simple analytical model and in a
Monte Carlo cascade model using dissociation cross sections obtained
from the model of Barnes and Swanson.  On the other hand, a critical
energy density model can explain the anomalous $J/\psi$ suppression,
suggesting the production of the quark-gluon plasma in central Pb+Pb
collisions at 158A GeV.

In conclusion, our re-assessment of these signatures provides
additional support to indicate that, in line with the summary of Heinz
and Jacobs \cite{Hei00}, the experimental data appear to be consistent
with the occurrence of the quark-gluon plasma in central Pb+Pb
collisions at 158A GeV.

\vskip 0.3cm The author would like to thank Drs.\ T.\ Awes, C.\ M.\
Ko, M.\ Murray, and N.\ Xu for helpful discussions.  The author would
like to thank Dr.\ H.\ Boggild for his permission to show the data of
the NA44 Collaboration (Fig.\ 3).  He would also like to thank his
collaborators, Drs.\ T.\ Barnes, E.\ Swanson, S.\ Sorensen, and B.\
H.\ Sa, for their collaboration to study $J/\psi$ dissociation cross
sections and $J/\psi$ suppression.

\end{document}